\documentclass[fleqn,12pt]{article}
\usepackage{amssymb}
\begin{document}
\baselineskip=0.6cm

\title{\bf On the Nonlocal Equations and Nonlocal Charges Associated with the Harry Dym Hierarchy}

\author{
J.C. Brunelli$^{a,}$\thanks{Supported by CNPq, Brazil. brunelli@fsc.ufsc.br\,.}\,,
G.A.T.F. da Costa$^{b,}$\thanks{gatcosta@mtm.ufsc.br\,.}\\
{\normalsize \it $^a$Departamento de F\'\i sica, CFM, Universidade Federal de Santa Catarina,}\\
\noalign{\vspace{-.1truecm}}
{\normalsize \it Campus Universit\'ario, Trindade, C.P. 476, CEP 88040-900, Florian\'opolis, SC, Brazil}\\
{\normalsize \it $^b$ Departamento de Matem\'atica, CFM,
Universidade Federal de Santa Catarina,}\\
\noalign{\vspace{-.1truecm}}
{\normalsize \it Campus Universit\'ario, Trindade, CEP 88040-900, Florian\'opolis, SC, Brazil}
 }

\maketitle

\begin{abstract}
A large class of nonlocal equations and nonlocal charges for the Harry Dym
hierarchy is exhibited. They are obtained from nonlocal Casimirs associated with its
bi-Hamiltonian structure. The Lax representation for some of these equations is also given.

\end{abstract}

\section{Introduction}
\label{sec:introduction}

 The following nonlinear partial differential equation
\begin{equation}
w_t=(w^{-1/2})_{xxx}\label{harrydym}\;,
\end{equation}
is known as the Harry Dym equation (see \cite{Hereman} for a
review). It can also be written in the following equivalent
forms
\begin{eqnarray}
v_t &=&{1\over 4}v^3 v_{xxx}\;,\nonumber\\
u_t&=&(u_{xx}^{-1/2})_x \;, \nonumber
\end{eqnarray}
for $v=-2^{1/3}w^{-1/2}$ and $u_{xx}=w$, respectively. This
equation was obtained by Harry Dym and Martin Kruskal as an
evolution equation solvable by a spectral problem based on the
string equation instead of the Schr\"odinger equation. This result
was reported in \cite{Kruskal} and rediscovered independently in
\cite{Sabatier,Shen}.
The Harry Dym equation share many of the properties typical of
the soliton equations. It is a completely integrable equation
which can be solved by the inverse scattering transform
\cite{Wadati1,Wadati2,Calogero}. It has a bi-Hamiltonian
structure and an infinite number of conservation laws and
infinitely many symmetries \cite{Magri,Olverbook}.

The nonlinear hyperbolic equation, which we call the Hunter-Zheng
equation,
\[
(u_t+u u_x)_{xx}={1\over2}(u_x^2)_x\;,
\]
or the nonlocal version
\begin{equation}
w_t=-(\partial^{-2}w)w_x-2(\partial^{-1}w)w\label{hunterzheng}
\end{equation}
for $u_{xx}=w$, has the same bi-Hamiltonian structure as the
Harry Dym equation. The complete integrability of
(\ref{hunterzheng}) was established in \cite{Hunter} as well its
connection with the Camassa-Holm equation \cite{Camassa}; the
former is the high-frequency limit of the latter. Due to the
presence of the antiderivative $\partial^{-1}$ the Hunter-Zheng
equation (\ref{hunterzheng}) is nonlocal.

As we will describe the Harry Dym equation (\ref{harrydym}) and
the Hunter-Zheng equation (\ref{hunterzheng}) belong to the same
hierarchy of flows, which we will call the Harry Dym hierarchy.
The Hunter-Zheng equation is a member of the positive order
equations in this hierarchy while the Harry Dym belongs to the
negative order equations. Hierarchies of negative order equations
were considered previously through the use of the negative powers
of the recursion operator \cite{Verosky,Andreev,Lou1}.
Usually, while the positive order equations are local the negative order ones are nonlocal.
For the Harry Dym hierarchy we have the opposite situation, all the positive order
equations  are nonlocal.

We  will show  the existence of two new hierarchies of integrable
nonlocal equations associated with the Harry Dym hierarchy
besides the Hunter-Zheng one. In fact the existence of two
nonlocal Casimirs operators implies some sort of degeneracy
for the positive order equations.

The paper is organized as follows. In Sec.~\ref{sec:bihamiltonian} we review the bi-Hamiltonian
formulation of integrable evolution equations emphasizing the role played by the Casimirs or
distinguished functionals of the Hamiltonian operators. We find one more nonlocal Casimir for the
modified Kortweg-de Vries equation. In Sec.~\ref{sec:HarryDym}  we obtain new
nonlocal equations as well nonlocal charges for the Harry Dym hierarchy. In Sec.~\ref{sec:Lax}
we discuss the Lax representation of these equations. The conclusions are given in
Sec.~\ref{sec:Conclusion}.

\section{Bi-Hamiltonian Systems and Casimirs}
\label{sec:bihamiltonian}

Central to our discussion is the concept of the bi-Hamiltonian
formulation of an integrable evolution equation
\cite{Magri,Olverbook}
\[
u_t=K_1[u]=\mathcal{D}_1 {\delta H_{2}\over\delta u}=\mathcal{D}_2
{\delta H_{1}\over\delta u}\;.
\]
Whenever $\mathcal{D}_1$    and  $\mathcal{D}_2$ are compatible
this implies the existence of an infinite hierarchy of higher
order commuting bi-Hamiltonian systems,
\begin{equation}
u_t^{(n)}=K_n[u]=\mathcal{D}_1 {\delta H_{n+1}\over\delta
u}=\mathcal{D}_2 {\delta H_{n}\over\delta u}\;,\quad\hbox{with $n\in
\mathbb{Z}$}\;,  \label{bihamiltonian}
\end{equation}
where the higher order conservation laws $H_n[u]$ are shared by
all members of the hierarchy. This hierarchy of equations can be
generated by the recursion operator
\[
R=\mathcal{D}_2\mathcal{D}_1^{-1}  \;,
\]
since
\[
K_{n+1}=RK_n\;.
\]
Also, using
\begin{equation}
{\delta H_{n+1}\over\delta u}=R^\dagger{\delta H_{n}\over\delta
u}\;,\label{recursioncharges}
\end{equation}
where $R^\dagger=\mathcal{D}_1^{-1}\mathcal{D}_2$  is the adjoint
of $R$, as a recursion scheme we can obtain the higher
Hamiltonians $H_n$.

We call any functional $H_C[u]$ a Casimir (or distinguished
functional) of the Hamiltonian operator $\mathcal{D}$ if
\[
\mathcal{D} {\delta H_{C}\over\delta u}=0\;.
\]
As a consequence any Hamiltonian system having  $\mathcal{D}$ as
a Hamiltonian operator,
\[
u_t=\mathcal{D} {\delta H\over\delta u}\;,
\]
has $H_C$ as a conserved charge. In fact
\begin{eqnarray}
\dot{H}_C=\{H_C,H\}&=&\int dx\int dy {\delta H_C\over\delta
u(x,t)}\{u(x,t),u(y,t)\} {\delta H\over\delta u(y,t)}\nonumber \\
&=&-\int  dx\left(\mathcal{D}{\delta H_C\over\delta u}\right)
{\delta H\over\delta u}=0\;. \nonumber
\end{eqnarray}
When one of the conservation laws in (\ref{bihamiltonian}) is a
Casimir, let us say of $\mathcal{D}_1$, the hierarchy of
equations stops except if the Hamiltonian operator
$\mathcal{D}_2$ has at least one Casimir.

The Kortweg-de Vries (KdV) equation
\[
u_t=u_{xxx}+3uu_x
\]
has the following series of conservation laws
\begin{eqnarray}
H_0&=&\int  dx \,u\;,\nonumber\\
H_1&=&\int  dx \,{1\over 2}u^2\;,\nonumber\\
 H_2&=&\int  dx \,{1\over 2}\left(u^3-u_x^2\right)\;,\nonumber\\
 &\vdots&  \nonumber
\end{eqnarray}
and the two compatible Hamiltonian operators
\begin{eqnarray}
\mathcal{D}_1&=&\partial\;,\nonumber\\
\mathcal{D}_2&=&\partial^3+u\partial+\partial u\;.\label{kdvstructures}
\end{eqnarray}
From (\ref{bihamiltonian}) we get
\begin{eqnarray}
u_t^{(0)}&=& \mathcal{D}_1 {\delta H_{1}\over\delta
u}=\mathcal{D}_2 {\delta H_{0}\over\delta u}=u_x\;,\nonumber\\
u_t^{(1)}&=& \mathcal{D}_1 {\delta H_{2}\over\delta
u}=\mathcal{D}_2 {\delta H_{1}\over\delta u}=u_{xxx}+3uu_x\;,\nonumber\\
u_t^{(2)}&=& \mathcal{D}_1 {\delta H_{3}\over\delta
u}=\mathcal{D}_2 {\delta H_{2}\over\delta u}=u_{xxxxx}+5uu_{xxx}+10u_xu_{xx}+{15\over2}u^2u_x\;,\nonumber\\
   &\vdots&\nonumber
\end{eqnarray}
and apparently we can not extend this recursion procedure for negatives
values of $n$ in  (\ref{bihamiltonian})    since $H_0$ is a
Casimir of $\mathcal{D}_1$ and $\mathcal{D}_2$ appears to have only trivial local
distinguished functionals (see Eq. \ref{kernelskdv}). Now,  the modified KdV equation (mKdV)
\[
u_t=u_{xxx}+{3\over2}u^2u_x
\]
has the following bi-Hamiltonian  form
\[
 u_t= \mathcal{D}_1 {\delta H_{2}\over\delta
u}=\mathcal{D}_2 {\delta H_{1}\over\delta u}\;,
\]
where
\begin{eqnarray}
\mathcal{D}_1&=&\partial\;,\nonumber\\
\mathcal{D}_2&=&\partial^3+\partial u\partial^{-1}u\partial\;,
\label{hamistrucmkdv}
\end{eqnarray}
and
\begin{eqnarray}
H_1&=&\int  dx \,{1\over 2}u^2\;,\nonumber\\
 H_2&=&\int  dx \,\left({1\over8}u^4-{1\over2}u_x^2\right)\;.\nonumber
\end{eqnarray}
Of course, $H_0=\int  dx\, u$ is the Casimir of $\mathcal{D}_1$,
however  $\mathcal{D}_2$ in (\ref{hamistrucmkdv}) admits a
nontrivial nonlocal Casimir \cite{Olver}
\[
H_C=\int  dx\, \cos(\partial_x^{-1} u)\;.
\]
Here we will define the skew-adjoint anti-derivative
$\partial^{-1}$ acting on functions $u$, which satisfy $u\to 0$ as
$|x|\to\infty$, by
\begin{equation}
(\partial_x^{-1} u)\equiv(\partial^{-1}u)(x)=\int_{-\infty}^{+\infty} dy\,\epsilon(x-y)u(y)\;, \label{dminus}
\end{equation}
where
\[
\epsilon(x-y)=\left\{\begin{array}{rl}
1/2&\hbox{ for }x>y\;,\\
-1/2&\hbox{ for } x<y\;.\end{array}\right.
\]
From now on we will omit the $x$ subscript in (\ref{dminus}). It
is easy to verify that for functions $A$ and $B$ of $u$ we have
the property
\[
\int  dx\, A(\partial^{-1}B)=-\int  dx(\partial^{-1}A)B\;.
\]
 Now
\[
{\delta H_{C}\over\delta
u}=\partial^{-1}\left(\sin(\partial^{-1}u)\right)
\]
and it is easy to show that $\mathcal{D}_2 ({\delta H_{C}/\delta
u})=0$. From (\ref{bihamiltonian}) we get, for $n=-1,-2,\dots\;,$ a hierarchy of negative order equations
\begin{eqnarray}
u_t^{(0)}&=& \mathcal{D}_1 {\delta H_{1}\over\delta
u}=\mathcal{D}_2 {\delta H_{0}\over\delta u}=u_x\;,\nonumber\\
u_t^{(-1)}&=& \mathcal{D}_1 {\delta H_{0}\over\delta
u}=\mathcal{D}_2 {\delta H_{-1}\over\delta u}=0\;,\nonumber\\
u_t^{(-2)}&=& \mathcal{D}_1 {\delta H_{-1}\over\delta
u}=\mathcal{D}_2 {\delta H_{-2}\over\delta u}=\sin(\partial^{-1}u)\;,\nonumber\\
   &\vdots& \label{negativemkdv}
\end{eqnarray}
where $H_{-1}\equiv H_C$. Introducing the potential function
$\psi_x=u$ the last equation in (\ref{negativemkdv}) is the
sine-Gordon equation
\[
\psi_{xt}=\sin\psi\;.
\]
Besides $H_0$ and $H_C$ we have found that the
Hamiltonian operator $\mathcal{D}_2$ also has the following nonlocal Casimir
\[
H_C'= \int  dx\, \sin(\partial^{-1}u)\left(\partial^{-1}\cos(\partial^{-1} u)\right)\;,
\]
which will generate another negative order hierarchy of equations.

Returning to the KdV equation if we set
\[
u=-2\psi^{-1}\psi_{xx}\;,
\]
the second Hamiltonian structure in (\ref{kdvstructures})  can be written as
\[
\mathcal{D}_2=\psi^{-2}\partial\psi^2\partial\psi^2\partial\psi^{-2}\;,
\]
and it follows that
\begin{equation}
\psi^2\;,\quad\psi^2(\partial^{-1}\psi^{-2})\;,\quad
\psi^2\partial^{-1}\left(\psi^{-2}(\partial^{-1}\psi^{-2})\right)\label{kernelskdv}
\end{equation}
are non trivial kernels of $\mathcal{D}_2$. We will not discuss
this system here but the Casimirs associated with
(\ref{kernelskdv}) will give rise to nonlocal KdV hierarchies of
equations (see \cite{Verosky,Andreev,Lou1}, and references
therein) and nonlocal charges.
Instead, we will perform this analysis in a systematic way for the Harry Dym hierarchy in the next section.

\section{The Harry Dym Hierarchy}
\label{sec:HarryDym}

The Harry Dym equation (\ref{harrydym}) is a completely
integrable bi-Hamiltonian system \cite{Magri,Olverbook}
\[
w_t =\mathcal{D}_1{\delta H_{-1}\over\delta
w}=\mathcal{D}_2{\delta H_{-2}\over\delta w}\;,
\]
where
\begin{eqnarray}
\mathcal{D}_1&=&\partial^3\nonumber\;,\nonumber\\
\mathcal{D}_2&=&w\partial+\partial w\;,
\label{harrydymstructures}
\end{eqnarray}
and
\begin{eqnarray}
H_{-1} &=&
\int  dx\,\left(2w^{1/2}\right)\nonumber\;,\\
H_{-2} &=& \int  dx\left({1\over8}w^{-5/2}w_x^2\right)\;.\nonumber
\end{eqnarray}
It is well known that $H_{-1}$ is a Casimir of  $\mathcal{D}_2$
and that
\begin{equation}
H_{0} =- \int  dx\,w \label{casimir}
\end{equation}
is a Casimir of  $\mathcal{D}_1$. So, we can consider equations
going ``up'' and ``down'' in the equation (\ref{bihamiltonian}).
However, we also have the following nonlocal Casimirs for $\mathcal{D}_1$
 \begin{eqnarray}
H_0^{(1)} &=&
\int  dx\,\left(\partial^{-1}w\right)\nonumber\;,\\
H_0^{(2)} &=& \int  dx\,\left(\partial^{-2}w\right)\;.\label{casimirnew}
\end{eqnarray}
In this way (\ref{bihamiltonian}) gets degenerated for $n>0$
 \begin{eqnarray}
 &\vdots&\nonumber\\
w_t^{(2,\alpha)}&=& \mathcal{D}_1 {\delta
H_{2}^{(\alpha)}\over\delta
w}=\mathcal{D}_2 {\delta H_{1}^{(\alpha)}\over\delta w}\;,\nonumber\\
w_t^{(1,\alpha)}&=& \mathcal{D}_1 {\delta
H_{1}^{(\alpha)}\over\delta
w}=\mathcal{D}_2 {\delta H_{0}^{(\alpha)}\over\delta w}\;,\nonumber\\
w_t^{(0)}&=& \mathcal{D}_1 {\delta H_{0}^{(\alpha)}\over\delta
w}=\mathcal{D}_2 {\delta H_{-1}\over\delta w}=0\;,\nonumber\\
w_t^{(-1)}&=& \mathcal{D}_1 {\delta H_{-1}\over\delta
w}=\mathcal{D}_2 {\delta H_{-2}\over\delta w}=(w^{-1/2})_{xxx}\;,\nonumber\\
   &\vdots& \label{harrydymhierarchy}
\end{eqnarray}
where $\alpha=0,1,2$ and $H_0^{(0)}=H_0$.

Using (\ref{recursioncharges})  and (\ref{harrydymhierarchy}) as a
recursion scheme we can obtain, after a straightforward but
tedious calculation, the first few Hamiltonian functionals and
flows for the Harry Dym hierarchy equations. For $n\le 0$ the
first conserved charges, some of them already calculated in
\cite{Hunter}, are
\begin{eqnarray}
H_0=H_{0}^{(0)}&=&\int  dx\,(-w)\nonumber\;,\\\noalign{\vskip
5pt}
H_{-1}&=&\int  dx\,2w^{1/2}\nonumber\;,\\\noalign{\vskip
5pt}
H_{-2}&=&\int  dx\,{1\over8}w^{-5/2}w_x^2\nonumber\;,\\\noalign{\vskip 5pt}
H_{-3}&=&\int  dx\,{1\over
16}\left({35\over16}w^{-11/2}w_x^4-w^{-7/2}w_{xx}^2\right)\nonumber\;,\\\noalign{\vskip
5pt}
H_{-4}&=&\int  dx\,{1\over
32}\left({5005\over128}w^{-17/2}w_x^6-{231\over8}w^{-13/2}w_x^2w_{xx}^2+
5w^{-11/2}w_{xx}^3\right.\nonumber\\
&&\left.\phantom{1\over2}+w^{-9/2}w_{xxx}^2\right)\nonumber\;,\\
\noalign{\vskip 5pt}
&\vdots&  \label{harrydymcharges-}
\end{eqnarray}
and the first flows are
\begin{eqnarray}
w_t^{(0)}&=&0\;,\nonumber\\\noalign{\vskip 5pt}
w_t^{(-1)}&=&(w^{-1/2})_{xxx}\;,\nonumber\\\noalign{\vskip 5pt}
w_t^{(-2)}&=&{1\over4}\left({5\over4}w^{-7/2}w_x^2-w^{-5/2}w_{xx}\right)_{xxx}
\;,\nonumber\\\noalign{\vskip 5pt}
w_t^{(-3)}&=&{1\over16}\left({1155\over32}w^{-13/2}w_x^4-{231\over4}w^{-11/2}w_x^2w_{xx}+
{21\over2}w^{-9/2}w_{xx}^2
\right.\nonumber\\
&&\left.+14w^{-9/2}w_{xxx}w_x-2w^{-7/2}w_{xxxx}\phantom{1\over2}\right)_{xxx}\;.\nonumber\\
&\vdots& \label{harrydymflows-}
\end{eqnarray}

For $n>0$ we have to consider the three cases $\alpha=0,1,2$
separately. So, for $\alpha=0$ we get the nonlocal conserved charges
\begin{eqnarray}
H_{1}^{(0)}&=&\int  dx\,{1\over
2}(\partial^{-1}w)^2\nonumber\;,\\\noalign{\vskip 5pt}
H_{2}^{(0)}&=&\int  dx\,{1\over
2}(\partial^{-2}w)(\partial^{-1}w)^2\nonumber\;,\\\noalign{\vskip
5pt} H_{3}^{(0)}&=&\int  dx\,\left[{1\over
4}(\partial^{-2}w)^2(\partial^{-1}w)^2+{1\over 8
}\left(\partial^{-1}(\partial^{-1}w)^2\right)^2\right]\nonumber\;,\\\noalign{\vskip
5pt} H_{4}^{(0)}&=&\int  dx\,\left[{1\over
12}(\partial^{-2}w)^3(\partial^{-1}w)^2 -{1\over
4}(\partial^{-1}w)^2\partial^{-2}\left((\partial^{-2}w)(\partial^{-1}w)^2\right)\right.\nonumber\\
&&\left.-{1\over 8
}(\partial^{-2}w)\left(\partial^{-1}(\partial^{-1}w)^2\right)^2\right]\;,\nonumber\\\noalign{\vskip
5pt} H_{5}^{(0)}&=&\int  dx\,\left[{1\over
48}(\partial^{-2}w)^4(\partial^{-1}w)^2 -{1\over
8}(\partial^{-2}w)^2(\partial^{-1}w)^2\left(\partial^{-2}(\partial^{-1}w)^2\right)\right.\nonumber\\
&&-{1\over
16}(\partial^{-2}w)^2\left(\partial^{-1}(\partial^{-1}w)^2\right)^2+{1\over
16}\left(\partial^{-2}(\partial^{-1}w)^2\right)^2(\partial^{-1}w)^2\nonumber\\
&&\left.+{1\over
16}(\partial^{-1}w)^2\left(\partial^{-2}\left(\partial^{-1}(\partial^{-1}w)^2\right)^2\right)+
{1\over8}\left(\partial^{-1}\left((\partial^{-1}w)\partial^{-1}(\partial^{-1}w)^2\right)\right)^2
\right]\;.\nonumber\\
&\vdots& \label{nonlocalh0}
\end{eqnarray}
and the first flows are
\begin{eqnarray}
w_t^{(1,0)}&=&-w_x\;,\nonumber\\\noalign{\vskip 5pt}
w_t^{(2,0)}&=&-(\partial^{-2}w)w_x-2(\partial^{-1}w)w\;,\nonumber\\\noalign{\vskip 5pt}
w_t^{(3,0)} &=&-{1\over 2}(\partial^{-2}w)^2
w_x-2(\partial^{-2}w)(\partial^{-1}w)w+{1\over2}\left(\partial^{-2}(\partial^{-1}w)^2\right)w_x
\nonumber\\
&&+w\left(\partial^{-1}(\partial^{-1}w)^2\right)\;,\nonumber\\ \noalign{\vskip 5pt}
&\vdots& \label{harrydymflows0+}
\end{eqnarray}
Note that the flow $w_t^{(2,0)}$ is the Hunter-Zheng equation
(\ref{hunterzheng}).

From (\ref{harrydymstructures}) we can construct the recursion
operator
\begin{equation}
R=
\mathcal{D}_2\mathcal{D}_1^{-1}=2w\partial^{-2}+w_x\partial^{-3}\;,\label{harrydymrecursion}
\end{equation}
and since
\[
(\partial w+w\partial)^{-1}={1\over2}w^{-1/2}\partial^{-1}w^{-1/2}
\]
we have
\[
R^{-1}={1\over 2}\,\partial^3 w^{-1/2} \partial^{-1}w^{-1/2}\;.
\]
The flows (\ref{harrydymflows-}) and (\ref{harrydymflows0+}) can
now be expressed as the action of powers of $R$ acting on the
seed equation $w_t^{(1,0)}=\mathcal{D}_2\left({\delta
H_0^{(0)}/\delta w}\right)=-w_x$
\begin{eqnarray}
w_t^{(n)} &=&R^{n-1}(-w_x)\;,\quad n=0,-1,-2,\dots\;,\label{iteration1}\\
w_t^{(n,0)} &=&R^{n-1}(-w_x)\;,\quad n=1,2,3,\dots\;.
\label{iteration2}
\end{eqnarray}
 To be able to perform the steps in the iteration given in (\ref{iteration1}-\ref{iteration2})
we must point out that for any function $f(x,t)$ we can use
instead of (\ref{dminus}) the representation
\cite{Sanders,Verosky,Andreev,Lou1}
\[
(\partial^{-1}f)(x,t)=\int_{-\infty}^{+\infty} dy\,\epsilon(x-y)f(y,t)+c(t)\;,
\]
where $c(t)$ is a function of $t$. Here we set $c=0$ for any $f$,
except $f=0$ where we use $c=2$. In this way
\[
w_t^{(0)}=R^{-1}(-w_x)=0
\]
and
\[
w_t^{(-1)}=R^{-2}(-w_x)={1\over2}\,\partial^3w^{-1/2}\partial^{-1}0=(w^{-1/2})_{xxx}\;,
\]
resulting in the Harry Dym equation.

For $n>0$ and $\alpha=1,2$ we have the seed equations
\begin{eqnarray}
w_t^{(1,1)} &=& \mathcal{D}_2{\delta H_{0}^{(1)}\over\delta
w}=-2w-xw_x\;,\nonumber\\
w_t^{(1,2)} &=& \mathcal{D}_2{\delta H_{0}^{(2)}\over\delta
w}=2xw+{x^2\over 2}w_x\;, \label{harrydymseed}
\end{eqnarray}
where we have used $(\partial^{-n}1)=x^n/n$, for $n>0$. The
respective flows follow from the analogue of (\ref{iteration2})
and they read for $\alpha=1$
\begin{eqnarray}
w_t^{(1,1)}&=&-2w-xw_x\;,\nonumber\\\noalign{\vskip 5pt}
w_t^{(2,1)}&=&-2wx(\partial^{-1}w)-w_x\left(\partial^{-1}(x(\partial^{-1}w))\right)
\;,\nonumber\\\noalign{\vskip
5pt} w_t^{(3,1)} &=&\left(2w+w_x\partial^{-1}\right)
\left({x\over2}\,\partial^{-1}(\partial^{-1}w)^2-(\partial^{-1}w)\partial^{-1}
(x(\partial^{-1}w))\right)
\;,\nonumber\\
&\vdots& \label{harrydymflows1+}
\end{eqnarray}
and for $\alpha=2$
\begin{eqnarray}
w_t^{(1,2)}&=&2xw+{x^2\over2}w_x\;,\nonumber\\\noalign{\vskip 5pt}
w_t^{(2,2)}&=&2w\left({1\over2}x^2(\partial^{-1}w)-(\partial^{-3}\omega)\right)+
w_x\,\partial^{-1}\left({1\over2}x^2(\partial^{-1}w)-(\partial^{-3}\omega)\right)
\;,\nonumber\\\noalign{\vskip 5pt} w_t^{(3,2)}
&=&\left(2w+w_x\partial^{-1}\right)
\left({1\over2}\,\partial^{-1}(\partial^{-2}w)^2+{1\over2}\partial^{-3}(\partial^{-2}w)^2
-(\partial^{-4}w)(\partial^{-1}w)\right.\nonumber\\
&&\left.-{x^2\over4}\partial^{-1}(\partial^{-1}w)^2+
{1\over2}(\partial^{-1}w)\partial^{-1}(x^2(\partial^{-1}w))\right)\;,\nonumber\\
&\vdots& \label{harrydymflows2+}
\end{eqnarray}
Again using (\ref{recursioncharges}) and (\ref{harrydymhierarchy}) recursively we obtain the
nonlocal conserved charges
\begin{eqnarray}
H_{1}^{(1)}&=&-{1\over2}\int  dx\,\partial^{-1}(\partial^{-1}w)^2\nonumber\;,\\\noalign{\vskip
5pt}
H_{2}^{(1)}&=&{1\over2}\int  dx\,\partial^{-1}\left((\partial^{-1}w)\partial^{-1}(\partial^{-1}w)^2\right)\nonumber\;,\\\noalign{\vskip
5pt}
H_{3}^{(1)}&=&-{1\over2}\int  dx\,\partial^{-1}\left[{1\over
4}\left(\partial^{-1}(\partial^{-1}w)^2\right)^2+(\partial^{-1}w)\partial^{-1}
\left((\partial^{-1}w)\partial^{-1}(\partial^{-1}w)^2\right)\right]\nonumber\;,\\\noalign{\vskip
5pt}
 &\vdots&\label{nonlocalh1}
\end{eqnarray}
and
\begin{eqnarray}
H_{1}^{(2)}&=&-{1\over2}\int  dx\,\left[(\partial^{-2}w)^2+\left(\partial^{-2}(\partial^{-1}w)^2
\right)\right]\nonumber\;,\\\noalign{\vskip 5pt}
H_{2}^{(2)}&=&{1\over2}\int  dx\,\left[{1\over3}(\partial^{-2}w)^3+
\partial^{-2}\left((\partial^{-1}w)\left(\partial^{-1}(\partial^{-1}w)^2\right)\right)
+(\partial^{-4}w)(\partial^{-1}w)^2\right]\nonumber\;,\\\noalign{\vskip 5pt}
H_{3}^{(2)}&=&-{1\over2}\int  dx\,\left[{1\over12}(\partial^{-1}w)^4
+{1\over4}\left(\partial^{-2}(\partial^{-1}w)^2\right)^2+
 {1\over4}\partial^{-2}\left(\partial^{-1}(\partial^{-1}w)^2\right)^2\right.\nonumber\\
 &&+\left.{1\over2}(\partial^{-1}w)^2\partial^{-2}(\partial^{-2}w)^2
 -(\partial^{-1}w)^2\partial^{-1}\left((\partial^{-4}w)(\partial^{-1}w)\right)\right.\nonumber\\
&&\left.+\partial^{-2}\left((\partial^{-1}w)\partial^{-1}\left((\partial^{-1}w)
 \left(\partial^{-1}(\partial^{-1}w)^2\right)\right)\right)\phantom{1\over4}\!\!\!\!
\right]\nonumber\;,\\\noalign{\vskip 5pt}
&\vdots&\label{nonlocalh2}
\end{eqnarray}
These hierarchies of integrable equations become extremely
nonlocal as we proceed further in the recursion. $w_t^{(2,1)}$ and
$w_t^{(2,2)}$ are equations of the Hunter-Zheng type. It is
interesting to note that the equations (\ref{harrydymflows0+}),
(\ref{harrydymflows1+}) and (\ref{harrydymflows2+}) are in the
positive direction (positive flows) of recursion while the Harry
Dym equations (\ref{harrydymflows-}) are in the negative
direction (negative flows). Of course this is due to the fact
that the recursion operator (\ref{harrydymrecursion}) is
completely nonlocal. That is to be compared with the usual
situation we have for the KdV recursion operator
$R=\partial^2+2u+u_x\partial^{-1}$.

After obtaining Equations (\ref{harrydymflows1+}) and
(\ref{harrydymflows2+}) we became aware of the papers \cite{Lou2}
and \cite{Fuchssteiner}  where these equations are given
implicitly in a recursion form. However, we have shown here their
origin from the Casimirs (\ref{casimir}) and (\ref{casimirnew}).
\section{Lax Pairs}
\label{sec:Lax}

The   equations in the Harry Dym hierarchy are integrable since
they are bi-Hamiltonian. Therefore we hope to find a Lax
representation for all of them. In fact, the Lax pair for the
Harry Dym equation (\ref{harrydym}) is given by
\cite{Hereman,Gesztesy,Konopelchenko}
\begin{eqnarray}
L&=&{1\over w}\partial^2\;,\label{harrydymlax}\\
B&=& -2w^{-3/2}\partial^3+{3\over
2}w^{-5/2}w_x\partial^2\;,\nonumber\\
 {\partial L\over\partial
t}&=&[B,L]\;.\label{harrydymlax0}
\end{eqnarray}
Calculating the square-root of $L$ (aided by a computer algebra
program) we obtain
\[
L^{1/2}=\beta\partial+a_0+a_1\partial^{-1}+a_2\partial^{-2}+a_3\partial^{-3}+a_4\partial^{-4}+
a_5\partial^{-5}+O(\partial^{-6})\;,
\]
where
\begin{eqnarray}
\beta &=& w^{-1/2}\;,\nonumber\\\noalign{\vskip 5pt}
a_0 &=& -{1\over2}\beta_x\;,\nonumber\\\noalign{\vskip 5pt}
a_1&=& {1\over 2^2}\beta_{xx}-{1\over
2^3}\beta_x^2\beta^{-1}\;,\nonumber\\\noalign{\vskip 5pt}
a_2 &=&-{1\over 2^3}\beta_{xxx}-{3\over
2^4}\beta_x^3\beta^{-2}+{3\over 2^3}\beta_x\beta_{xx}\beta^{-1}\;,\nonumber\\\noalign{\vskip 5pt}
a_3 &=& {1\over 2^4}\beta_{xxxx}-{3\over
2^3}\beta_x\beta_{xx}\beta^{-1}+{37\over
2^5}\beta_x^2\beta_{xx}\beta^{-2} -{61\over
2^7}\beta_x^{4}\beta^{-3}-{13\over
2^5}\beta_{xx}^2\beta^{-1}\;,\nonumber\\\noalign{\vskip 5pt}
a_4&=& -{1\over 2^5}\beta_{(5)}+{5\cdot7\over
2^5}\beta_{xx}\beta_{xxx}\beta^{-1}+{5\over
2^4}\beta_x\beta_{xxxx}\beta^{-1} -{3\cdot5\cdot7\over
2^6}\beta_x^{2}\beta_{xxx}\beta^{-2}\nonumber\\
&&-{3\cdot5\cdot13\over 2^6}\beta_x\beta_{xx}^2\beta^{-2}+
{3^2\cdot5\cdot7\over2^6}\beta_x^3\beta_{xx}\beta^{-3}-
{3\cdot5\cdot29\over 2^8}\beta_x^5\beta^{-4}\;,\nonumber\\\noalign{\vskip 5pt}
a_5&=& {1\over 2^6}\beta_{(6)}-{7\cdot17\over
2^7}\beta_{xxx}^2\beta^{-1}-{19\over
2^4}\beta_{xx}\beta_{xxxx}\beta^{-1} +{43\over
2^2}\beta_{xx}\beta_{xxx}\beta_x\beta^{-2}\nonumber\\
&&-{3\cdot5\over 2^6}\beta_x\beta_{(5)}\beta^{-1}+
{241\over2^7}\beta_x^2\beta_{xxxx}\beta^{-2}-
{569\over 2^6}\beta_x^3\beta_{xxx}\beta^{-3}+{413\over 2^7}\beta_{xx}^3\beta^{-2}\nonumber\\
&&-{3\cdot 1973\over 2^8}\beta_x^2\beta_{xx}^2\beta^{-3}+{3\cdot4493\over2^9}\beta_x^4\beta_{xx}\beta^{-4}
-{7\cdot17\cdot 67\over 2^{10}}\beta_x^6\beta^{-5}\;,\nonumber
\end{eqnarray}
where $\beta_{(n)}=d^n \beta/ dx^n$. Now it can be easily recognized that
\[
B=-2\left(L^{3/2}\right)_{\ge2}\;,
\]
and $\left(\ \right)_{\ge2}$ stands for the differential part of the pseudodifferential
operator with terms $\partial^n$, $n\ge2$. In this way (\ref{harrydymlax0})
assumes the nonstandard Lax representation
\begin{equation}
{\partial L\over\partial t}=-2\left[\left(L^{3/2}\right)_{\ge2},L\right]\;.\label{harrydymlax1}
\end{equation}
Similarly the whole negative hierarchy of equations (\ref{harrydymflows-}) can be obtained from
\begin{equation}
{\partial L\over\partial
t}=-2^n\left[\left(L^{{2n+1\over2}}\right)_{\ge2},L\right]\;,\quad n=0,1,2,\dots\;.\label{negativelax}
\end{equation}
The charges (\ref{harrydymcharges-}) (except by multiplicative constant factors) follows from
\begin{equation}
H_{-(n+1)}=\hbox{Tr} L^{2n-1\over 2}\;,\quad n=1,2,3,\dots\;,\label{negativehs}
\end{equation}
where ``$\hbox{Tr}$'' is the usual Adler trace \cite{Adler}. We have used (\ref{negativelax})
and (\ref{negativehs}) to perform a check of the equations (\ref{harrydymflows-}) and
charges (\ref{harrydymcharges-}), respectively.

Through ``gauge transformations'' \cite{Aratyn} the Lax
representation for the Harry Dym hierarchy (\ref{harrydymlax0})
and (\ref{harrydymlax1}) can be brought in other forms. For
instance, the Harry Dym equation also follows from
\[
L'=\partial\,L\,\partial^{-1}\;,
\]
with the nonstandard Lax representation
\begin{equation}
{\partial L'\over\partial
t}=-2\left[\left({L'}^{3/2}\right)_{\ge1},L'\right]\;, \label{nonstandardlax}
\end{equation}
or even from a standard Lax representation
\[
{\partial L''\over\partial
t}=-2\left[\left({L''}^{3/2}\right)_{\ge0},L''\right]\;,\nonumber
\]
with
\[
L''=w^{1/2}\,L\,w^{-1/2}\;. \nonumber
\]

So, for the negative flows of the Harry Dym hierarchy we have a complete Lax representation.
However, for the positive flows the Lax representation picture is not so complete.
It is easy to check that the Lax operator (\ref{harrydymlax}) with
\begin{equation}
 {\partial L\over\partial
t}=-2[B^{(i,\alpha)},L]\;,\quad i=1,2,3\;,\quad\alpha=0,1,2\;,\label{positivelax}
\end{equation}
yields the first equations $w^{(1,\alpha)}$, for the positive flows (\ref{harrydymflows0+}),
(\ref{harrydymflows1+}) and (\ref{harrydymflows2+}), if we choose
\begin{eqnarray}
B^{(1,0)}&=&{1\over 2}\partial\;,\nonumber\\
B^{(1,1)}&=&-{1\over4}+{1\over 2}x\partial\;,\nonumber\\
B^{(1,2)}&=&{1\over4}x-{1\over 4}x^2\partial\;, \nonumber
\end{eqnarray}
respectively. For the  equations $w^{(2,\alpha)}$ in (\ref{harrydymflows0+}),
(\ref{harrydymflows1+}) and (\ref{harrydymflows2+}) we have obtained, for $i=2$ in (\ref{positivelax}),
the operators
\begin{eqnarray}
B^{(2,0)}&=&{1\over 4}(\partial^{-2}w)\partial+{1\over4}\partial^{-1}(\partial^{-2}w)\partial^2\;,\nonumber\\
B^{(2,1)}&=&{1\over4}\left(\partial^{-1}(x\partial^{-1}w)\right)\partial +
{1\over4}\partial^{-1}\left(\partial^{-1}(x\partial^{-1}w)\right)\partial^2-
{1\over4}\partial^{-1}(\partial^{-1}w)\;,\nonumber\\
B^{(2,2)}&=&{1\over4}\left(\partial^{-1}\left({1\over2}x^2\partial^{-1}w-\partial^{-3}w\right)\right)\partial
+{1\over4}\partial^{-1}\left(\partial^{-1}\left({1\over2}x^2\partial^{-1}w-\partial^{-3}w\right)
\right)\partial^2\nonumber\\
&&-{1\over4}\partial^{-1}x(\partial^{-1}w)+{1\over4}\partial^{-1}(\partial^{-2}w)+{1\over8}\partial^{-2}xw \;.\nonumber
\end{eqnarray}
In fact $B^{(2,0)}$ was first obtained in \cite{Hunter} but in
the nonstandard Lax representation (\ref{nonstandardlax}). An
interesting question is how, if possible at all, $B^{(i,\alpha)}$
in (\ref{positivelax}) and the nonlocal charges $H_n^{(\alpha)}$
in (\ref{nonlocalh0}), (\ref{nonlocalh1}) and (\ref{nonlocalh2})
can be obtained from the same Lax operator (\ref{harrydymlax}),
i.e., what are the analogues of equations (\ref{negativelax}) and
(\ref{negativehs}) for the positive Harry Dym flows? In the
literature we can find Lax representations for equations obtained
by the inverse recursion operator \cite{Zhou,Qiao}, such as of the Harry Dym type.
However, these Lax representations are not given only in terms of $L$ and use some
sort of ansatz. Also, the authors do not try to obtain the relation between
the nonlocal charges and the Lax operator. These intriguing points are under investigation and will
appear elsewhere.

\section{Conclusion}
\label{sec:Conclusion}

We have given a unified picture of the Harry Dym hierarchy of equations which includes
local as well a series of three nonlocal hierarchies of equations. We have shown, using the
bi-Hamiltonian formulation of integrable models,  how the nonlocal Casimirs
leads to these nonlocal equations and also to three series of nonlocal charges.
Some of the nonlocal equations and nonlocal charges obtained in this way are new. This procedure can also be
applied for the KdV and mKdV equations since both equations, accordingly with the discussion in
Sec.~\ref{sec:bihamiltonian},  also have three Casimirs associated with the third order
Hamiltonian operator $\mathcal{D}_2$. We believe that the treatment given here
for the Harry Dym hierarchy unifies, within the bihamiltonian formulation of the integrable models,
some of the results scattered in the literature . We have also tried to understand
these nonlocal equations and charges from a Lax representation. Even though we have found explicitly Lax
pairs for some of the positive flows a unique Lax representation is still missing.

\end{document}